\documentclass[aps,preprint,amsmath]{revtex4}

\begin{document}
\draft
\title{ Optically induced spin polarization of an electric current through a quantum dot}
\author{Anatoly Yu. Smirnov }
\email{anatoly@dwavesys.com} \affiliation{ D-Wave Systems Inc.,
320-1985 West Broadway \\
 Vancouver, B.C. V6J 4Y3, Canada   }
\author{ Lev G. Mourokh}
\email{lmurokh@stevens.edu} \affiliation{ Department of Physics
and Engineering Physics \\ Stevens Institute of Technology,
Hoboken, NJ 07030, USA}
\date{\today }

\begin{abstract}
{We examine electron transport through semiconductor quantum dot
subject to a continuous circularly polarized optical irradiation
resonant to the electron - heavy hole transition. Electrons having
certain spin polarization experience Rabi oscillation and their
energy levels are shifted by the Rabi frequency. Correspondingly,
the equilibrium chemical potential of the leads and the
lead-to-lead bias voltage can be adjusted so only electrons with
spin-up polarization or only electrons with spin-down polarization
contribute to the current. The temperature dependence of the spin
polarization of the current is also discussed. }
\end{abstract}

 \maketitle

Recently, spin electronics, so-called {\it spintronics}, has
become a focal point of intense research interest (see, for
example, review \cite{Zutic1} and references therein). Electronic
devices employing the additional spin degree of freedom, as well
as purely spin-based devices, can increase the speed of
performance, consume less power, and provide novel modes of
operation in comparison with conventional counterparts. Creation,
control, and transmission of spin polarization constitute the
principal tasks of spintronics. To polarize electron and nuclear
spins, one can apply, for example, an external magnetic field (or put spin particles in contact with ferromagnetic materials) \cite{ClarkFeher,Moodera1}.
 Other possibilities for
creation of the spin polarization include optical orientation of
electron spins by a circularly polarized light \cite{OptPump1} and
polarized spin injection from the ferromagnetic/ normal
semiconductor junction \cite{Chye1}.

In the present paper we propose to combine the optical spin
polarization technique with electron transport through
semiconductor quantum dots to produce a spin polarization of the
electron current. Creating electron-hole pair with an electron
having certain spin projection, we prevent the tunneling of
electrons with the same spin projection to the quantum dot,
breaking the spin symmetry of the current. Electron spin
polarization in the dot can be achieved by the circular polarized
light. In particular, the photons with the positive helicity
create heavy holes with spin 3/2 and electrons with spin -1/2,
whereas the photons having negative helicity create heavy holes
with spin -3/2 and electrons with spin 1/2. For the self-assembled
quantum dots (for example, InGaAs/GaAs), the lattice mismatch
leads to the removal of the degeneracy between heavy and light
hole energy levels, and the electron-heavy hole pairs can be
created resonantly. Under resonant continuous optical irradiation,
electron-hole pairs (with certain electron spin projection)
exhibit Rabi oscillations \cite{RabiEx}, and the energy level for
the electrons having this spin projection is shifted by $\pm
\Omega_R$, where $\Omega_R$ is the Rabi frequency. We show that in
this case the conductance peaks occur not when the chemical
potential of the leads is in the resonance with the energy of the
quantum dot level, $E_c$, but when it is in the resonance with the
energy $E_c\pm \Omega_R$  ($\hbar =1$). However, for electrons
having the spin projection which is not affected by the light, the
conductance peak position remains the same. Accordingly, for
appropriate values of the Rabi frequency, the equilibrium chemical
potential of the leads, and the lead-to-lead voltage bias, the
spin polarization of the electron current through the quantum dot
can be achieved and, moreover, the direction of the polarization
can be controlled by these parameters.

We examine a self-assembled InGaAs/GaAs semiconductor quantum dot
which is electrically coupled to two leads (similar structure was
examined experimentally, for example, in Ref. \cite{Warbur}) and
is subject to a continuous circularly polarized light having
frequency resonant to the electron-heavy hole transition and
negative helicity. In this case, photons create an electron-hole
pairs in the dot with electrons and holes having spin projections
1/2 and -3/2, respectively. Electrons with both spin projections
can tunnel from the leads to the dot with spin-independent
tunneling rates, whereas the hole tunneling is assumed to be
suppressed.

The Hamiltonian of the system has the form
\begin{eqnarray}
H = E_c(a_1^+a_1 + a_2^+a_2) + E_v d^+d + \sum_{\alpha ,k, \sigma}E_{\alpha k}
c^+_{\alpha k,\sigma}c_{\alpha k,\sigma}- \nonumber\\
F(t)(a_1^+d^+ + da_1) - \sum_{\alpha ,k, \sigma}(T_{\alpha
k}c_{\alpha k,\sigma}^+a_{\sigma} + T_{\alpha
k}^*a_{\sigma}^+c_{\alpha k,\sigma}),
\end{eqnarray}
with  $\alpha = L,R$ and spin index $\sigma = 1,2.$  Operators
$c_{\alpha k,\sigma}$ are related to electrons in leads, whereas
$a_1,a_2$ describe an electron in the dot with spin projections
$\pm1/2$ as $a_1 = a_{1/2}$ and $a_2 = a_{-1/2}.$ The holes have
only one projection, so that $ d= d_{-3/2}. $ The monochromatic
electromagnetic field is given by $F(t) = F e^{i\omega_0 t} + F^*
e^{-i\omega_0 t}.$ The Coulomb interaction between electrons and
holes is neglected. From this Hamiltonian we derive the equations
of motion for the electron/hole amplitudes as
\begin{subequations}
\begin{equation}
i \dot{a_1} = E_c a_1 - F(t) d^+ - \sum_{\alpha ,k}T^*_{\alpha
k}c_{\alpha k,1},
\end{equation}
\begin{equation}
i \dot{a_2} = E_c a_2 - \sum_{\alpha ,k}T^*_{\alpha k}c_{\alpha
k,2},
\end{equation}
\begin{equation}
i \dot{d}^+ = -E_vd^+ - F(t) a_1,
\end{equation}
and
\begin{equation}
i\dot{c}_{\alpha k,\sigma}=E_{\alpha k}c_{\alpha k,\sigma} -
T_{\alpha k} a_{\sigma}.
\end{equation}
\end{subequations}
Introducing new electron/hole variables, $A_{\sigma} = a_{\sigma}
e^{iE_ct}$ and $D = d e^{iE_vt}, $, we rewrite Eq. (2) as
\begin{subequations}
\begin{equation}
i \dot{A_1} = F(t) e^{i(E_c+E_v)t} D^+ - \sum_{\alpha ,k}
T^*_{\alpha k} e^{iE_c t} c_{\alpha k,1},
\end{equation}
\begin{equation}
i \dot{A_2} =  - \sum_{\alpha ,k}T^*_{\alpha k} e^{i E_c t}
c_{\alpha k,2},
\end{equation}
\begin{equation}
i \dot{D}^+ =  - F(t) e^{-i(E_c+E_v)t} A_1,
\end{equation}
and
\begin{equation}
i \dot{c}_{\alpha k,\sigma}=E_{\alpha k}c_{\alpha k,\sigma} -
T_{\alpha k} e^{- i E_c t} A_{\sigma}.
\end{equation}
\end{subequations}
For the resonant electromagnetic field with the frequency
$\omega_0 = E_c + E_v$, the rotating wave approximation (RWA) can
be applied and, accordingly, $ F(t) e^{i(E_c+E_v)t} \simeq F^*,
F(t) e^{-i(E_c+E_v)t} \simeq F.$ Correspondingly, the equation for
the amplitude $A_1$ has the form
\begin{equation}
\left[\left(i \frac{d}{dt}\right)^2 - |F|^2 \right] A_1 = - i
\frac{d}{dt}\sum_{\alpha k} T^*_{\alpha k} e^{iE_c t} c_{\alpha
k,1},
\end{equation}
where the amplitude of the electromagnetic field, $|F|$, denotes
the Rabi frequency: $\Omega_R = |F|.$ The response of the
leads is described by the relation \cite{SHM}
\begin{equation}
c_{\alpha k,\sigma}(t) = c_{\alpha k,\sigma}^{(0)}(t) - T_{\alpha
k} \int dt_1 g_{\alpha k}^r(t,t_1) e^{-iE_c t_1} A_{\sigma}(t_1),
\end{equation}
where
\begin{subequations}
\begin{equation}
g_{k\alpha}^r(t,t_1)= - i e^{-iE_{k\alpha}(t-t_1)} \theta(t-t_1)
\end{equation}
and
\begin{equation}
\langle c_{k\alpha,\sigma}^{(0)}(t_1)^+ c_{k\alpha,\sigma}^{(0)}(t)\rangle = f_{\alpha}(E_{k\alpha})
e^{-iE_{k\alpha}(t-t_1)}
\end{equation}
\end{subequations}
are the Green functions of electrons in the leads,
$\theta(t-t_1)$ is the unit Heaviside function, and $f_{\alpha}(E)
= \left[\exp\left(\frac{E-\mu_{\alpha}}{T}\right)+1\right]^{-1}$
is the Fermi distribution with temperature $T$. Here, $\mu_L = \mu
+ eV/2$ and $\mu_R = \mu - eV/2,$ where $\mu$ is the equilibrium
chemical potential of the leads and $V$ is the voltage bias
applied to the leads.

Substituting Eq. (5) into Eq. (4), we obtain
\begin{eqnarray}
\left[\left(i \frac{d}{dt}\right)^2 - \Omega_R^2 \right] A_1 -
i\frac{d}{dt}\sum_{\alpha ,k}  |T_{\alpha k}|^2 \int dt_1
g_{\alpha k}^r(t,t_1) e^{i E_c(t-t_1)} A_1(t_1) = \nonumber\\
- i \frac{d}{dt}\sum_{\alpha ,k} T^*_{\alpha k} e^{iE_c t}
c_{\alpha k,1}^{(0)}.
\end{eqnarray}
The solution of this equation can be written as
\begin{equation}
A_1(t) = - i \frac{d}{dt} \int dt' G_1^r(t-t') \sum_{\alpha ,k}
T^*_{\alpha k} e^{iE_c t'} c_{\alpha k,1}^{(0)}(t'),
\end{equation}
where the Fourier transform of the retarded Green function $
G_1^r(t-t') $ is given by
\begin{equation}
G_1^r(\omega ) = \frac{1}{ \omega^2 - \Omega_R^2 - \omega
\sum_{\alpha ,k} |T_{\alpha k}|^2 g_{\alpha k}^r (\omega + E_c) },
\end{equation}
with $ g_{\alpha k}^r(\omega) = 1/(\omega - E_{\alpha k}) - i \pi
\delta (\omega - E_{\alpha k}).$ Broadenings of the electron level
in the dot due to its connection to the leads is described by the
damping coefficients $\Gamma_{\alpha}(\omega ) = 2\pi \sum_k
|T_{k\alpha}|^2 \delta (\omega - E_{k\alpha }),$ whereas the real
part of the function $ g_{\alpha k}^r (\omega + E_c) $ produces
insignificant  corrections to the Rabi frequency $\Omega_R.$
With a notation for  the average linewidth of the electron level, $\Gamma(\omega) =(1/2) [\Gamma_L(\omega) +
\Gamma_R(\omega)]$,
we obtain the expression for the Green function $G_1^r(\omega)$ as
\begin{equation}
G_1^r(\omega ) = \frac{1}{ \omega^2 - \Omega_R^2 + i \omega \Gamma(\omega + E_c) }.
\end{equation}

The electron current through the quantum dot is given by
\begin{equation}
I_{\alpha,\sigma}=\frac{d}{dt}\sum_k \langle c_{\alpha k,
\sigma}^+(t) c_{\alpha k, \sigma}(t)\rangle = i \sum_k T_{\alpha
k} e^{-iE_c t} \langle c_{\alpha k,\sigma}^+ A_{\sigma}  \rangle
+ h.c. ,
\end{equation}
It should be noted that the total current has two components,
$I_1$ with spin projection +1/2 and $I_2$ with spin projection
-1/2. Substituting Eq.(5) into Eq. (11), we obtain
 \begin{eqnarray}
I_{\alpha,\sigma}= i \sum_k T_{\alpha k} e^{-iE_c t}
\langle (c_{\alpha k,\sigma}^{(0)})^+ A_{\sigma}  \rangle - \nonumber\\
i \sum_k |T_{\alpha k}|^2 \int dt' [ g_{\alpha k}^r(t,t')]^+
e^{-iE_c(t-t')} \langle A_{\sigma}(t')^+ A_{\sigma}(t)\rangle +
h.c.
\end{eqnarray}
The first term in the right-hand side of Eq. (12) for the spin-up
polarized current $(\sigma = 1)$ can be evaluated as
\begin{equation}
i \sum_k T_{\alpha k} e^{-iE_c t} \langle (c_{\alpha
k,1}^{(0)})^+ A_{1}  \rangle +h.c. = 2 \int \frac{d\omega}{2\pi}
(\omega - E_c) Im [G_1^r(\omega - E_c)] f_{\alpha}(\omega
)\Gamma_{\alpha}(\omega ).
\end{equation}
It follows from Eq.(8) that the correlator $\langle A_{1}(t')^+
A_{1}(t)\rangle$ has the form $(\alpha, \beta = L,R)$
\begin{equation}
\langle A_{1}(t')^+ A_{1}(t)\rangle = \int \frac{d\omega}{2\pi}
e^{-i\omega (t-t')}\omega^2 |G_1^r(\omega)|^2 \sum_{\beta}
\Gamma_{\beta}(\omega + E_c) f_{\beta}(\omega+ E_c),
\end{equation}
and, consequently, the second term of the right-hand side of Eq.
(12) for the spin-up polarized current can be written as
\begin{eqnarray}
-i \sum_k |T_{\alpha k}|^2 \int dt' [ g_{\alpha k}^r(t,t')]^+ e^{-iE_c(t-t')} \langle A_{1}(t')^+ A_{1}(t)\rangle + h.c. = \nonumber\\
\int \frac{d\omega}{2\pi} (\omega - E_c)^2 |G_1^r(\omega - E_c)|^2 \sum_{\beta}\Gamma_{\alpha}(\omega)\Gamma_{\beta}(\omega) f_{\beta}(\omega ).
\end{eqnarray}
The total spin-up polarized current, $I_{\alpha,1}$ is equal to
the sum of the terms, Eqs. (13) and (15), and for the steady-state
regime, the current carrying the spin $+1/2$ has the form
\begin{equation}
I_1 = I_{L,1} = - I_{R,1} =  \int \frac{d\omega}{2\pi}
\frac{(\omega - E_c)^2  [ f_L(\omega ) - f_R(\omega) ] }{ [
(\omega - E_c)^2 - \Omega_R^2 ]^2 + [(\omega - E_c)
\Gamma(\omega)]^2 } \Gamma_L(\omega ) \Gamma_R (\omega).
\end{equation}
The current with the spin polarization -1/2 (which is not affected
by the optical irradiation) can be determined using the same
procedure as
\begin{equation}
I_2 =  \int \frac{d\omega}{2\pi} \frac{  [ f_L(\omega ) -
f_R(\omega) ] } {  (\omega - E_c)^2 + \Gamma^2(\omega) }
\Gamma_L(\omega ) \Gamma_R (\omega).
\end{equation}

Components of a zero-temperature conductance of the system,
$G_{\pm 1/2}  = \frac{d}{dV}(e I_{1,2}),$ related to the different
spin projections, exhibit the resonant behavior as functions of
the equilibrium chemical potential of the leads,  $\mu$,   ( or
the gate voltage applied to the dot) as
\begin{equation}
G_{+1/2} = \frac{e^2}{h} \times \frac{ (\mu - E_c)^2 \Gamma^2 } {
[(\mu - E_c)^2 - \Omega_R^2]^2 + (\mu - E_c)^2 \Gamma^2 },
\end{equation}
\begin{equation}
G_{-1/2} = \frac{e^2}{h} \times  \frac{  \Gamma^2 } { (\mu -
E_c)^2  + \Gamma^2 },
\end{equation}
where we assume symmetric coupling to the leads with $\Gamma =
\Gamma_{L/R} = \Gamma_{L/R}(\omega = \mu).$ This resonant behavior
is illustrated by Figure 1 for the Rabi frequency $\Omega_R =
2\Gamma$. One can see from this figure that the conductance peaks
for the electrons having spin projection 1/2 (affected by the
optical irradiation) are shifted from the equilibrium resonant
condition ($\mu =E_c$) by $\pm\Omega_R$. Furthermore, it is
evident that with variation such parameters as the equilibrium
chemical potential of the leads, the Rabi frequency, and the
lead-to-lead voltage bias, regimes with different spin
polarization of the current through the quantum dot can be
achieved.

To describe the spin polarization of the current and its
dependence on the system parameters, we introduce the polarization
coefficient as
\begin{equation}
P = {I_1-I_2\over I_1+I_2}.
\end{equation}
The lead-to-lead voltage dependencies of the total current and the
polarization coefficient are shown in Figure 2 for $\mu -E_c =
5\Gamma$, $\Omega_R = 20\Gamma$, and $T = 0.43\Gamma$. It is
evident from this figure that there are several steps in the
current-voltage characteristics. The first one occurs when the
chemical potential of the right lead, $\mu_R=\mu -eV/2$, passes
through the electron level with the energy $E_c$ and this level
becomes conductive. The electrons in this level have spin
projection -1/2 and the current is strongly polarized. The second
step corresponds to the voltage at which the chemical potential of
the left lead, $\mu_L=\mu + eV/2$, passes through the electron
level with the energy $E_c+\Omega_R$ and electrons from this level
having spin projection +1/2 start to contribute to the current.
Accordingly, current becomes only partially polarized. Finally,
the third step occurs when the chemical potential of the right
lead passes through the electron level with the energy
$E_c-\Omega_R$ and the spin polarization of the current is fully
compensated.

The temperature dependence of the polarization coefficient is
presented in Figure 3 for small applied lead-to-lead bias and two
values of the equilibrium chemical potential of the leads
associated with the resonances with electron levels having
different spin polarization. One can see in this Figure that at
temperatures comparable to the Rabi frequency, the spin
polarization of the current vanishes.

Finally, we consider the conditions necessary to observe the
calculated effects in experiment. If we assume the lead-dot
coupling constant to be $\Gamma = 20\mu eV$, than $ \Omega_R = 20
\Gamma = 400 \mu eV$, which corresponds to the excitation density
of $350 kW/cm^2$, and temperature used for Figure 2 is $T =
0.43\Gamma = 100mK$.

In conclusion, we have shown that the spin polarization of the
electron current through the quantum dot can be achieved, if the
dot is irradiated continuously by the resonant circularly
polarized light. In this case electron-hole pairs with electrons
having certain spin polarization experience Rabi oscillations and
the energetic level of these electrons is shifted by $\pm
\Omega_R$. With appropriate choice of the equilibrium chemical
potential of the leads and the lead-to-lead bias voltage, the
current through the dot is spin-polarized and the direction of
this polarization can be controlled.

Authors would like to thank A. O. Govorov for helpful discussions.
L. G. M. gratefully acknowledges support from the Department of
Defense, Grant No DAAD 19-01-1-0592. A. Yu. S. is thankful to A.
M. Zagoskin for critical reading of manuscript.

\newpage

\begin{center}
{\large Figures}
\end{center}

Figure 1. Zero-temperature conductances for the electrons having
different spin projections as a function of the equilibrium
chemical potential of the leads.

Figure 2. The total current and the current polarization as
functions of the applied lead-to-lead bias.

Figure 3. Temperature dependence of the current polarization.

\end{document}